\documentclass[prl,twocolumn,graphicx,amssymb,floatfix]{revtex4}

\usepackage{graphicx}
\begin{document}

{\bf Comment on ``Protocol for Direct Counterfactual Quantum Communication". }

In a recent letter  Salih et al. \cite{Zub13} claimed that ``in the ideal asymptotic limit, information can be transferred
between Alice and Bob without any physical particles traveling between them''.  This conclusion was based on a naive classical approach to the past of the photons: ``the photon could not have been in the transmission channel because it could not pass through it''. I will argue that actual measurement of the presence of the photon in the transmission channel will not support this claim.

Salih et al. build their protocol on the basis of the interaction-free measurements \cite{EV93} which were implemented as counterfactual computation \cite{Joz98} and counterfactual cryptography \cite{Noh09}. In all these scenarios, blocking the wave packet of a photon spoils destructive interference and information about the blocking is obtained without the photon being near the blockade \cite{Va03}. If the blockade is absent, these protocols cease to be counterfactual.  Salih et al., similarly to Hosten et al. \cite{Ho06} construct a protocol which is apparently counterfactual in both cases, when the blockade is present and when it is not. I have argued before \cite{Va07} that Hosten et al. method is not counterfactual for the null outcome  and, similarly, I claim that Salih et al. protocol is not counterfactual for  the values of the information bit corresponding to the absence of the blockade.

Salih et al. are correct that the branch of the wave function of the photon reaching detector $D_1$ does not pass through the communication channel.  However, from this does not follow that the photon was not there.  Both forward and backward evolving wave functions are present  in the communication channel, see Fig. 1., and I argue that in such a case we should say that the photon was there \cite{Va13}. Salih et al. are mistaken in saying ``the probability of finding a signal photon in the transmission channel is nearly zero''. Given a click at $D_1$,  the probability for finding the photon by  a nondemolition measurement of the projection operator on the transmission channel is one! Indeed, performing such a measurement and not finding the photon is impossible, since this is equivalent to blocking the channel, in which case $D_2$ has to click. The strong measurement of the projection completely changes the interference pattern, so the  relevant question is the outcome of a weak measurement of the presence of the photon in the communication channel. But when the strong measurement outcome is certain, the weak measurement yields the same result \cite{AV91}.  ``Logic 0'' case requires the photon to be present in the transmission channel.

\begin{figure}
\includegraphics[width=0.49\textwidth]{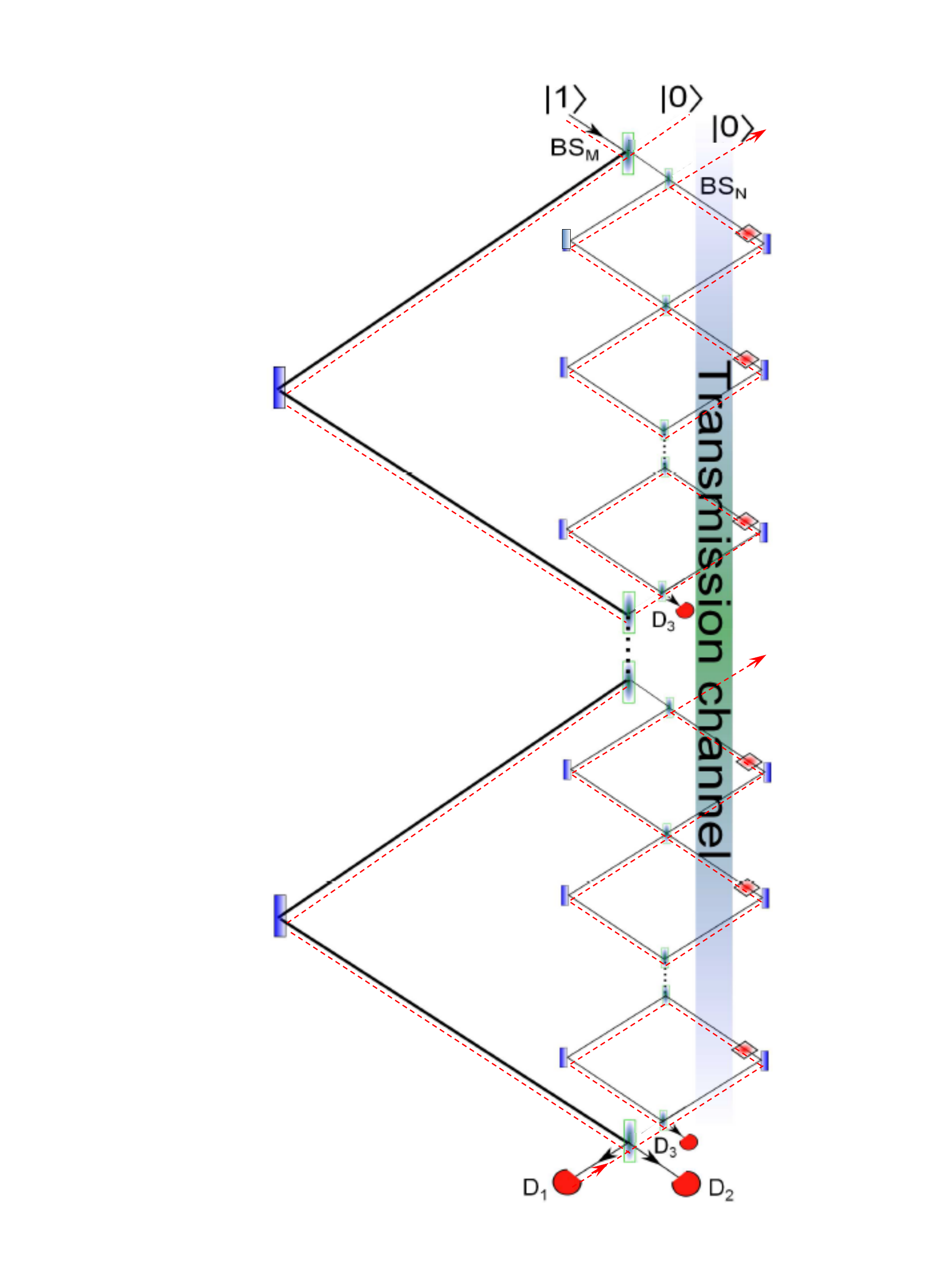}
\caption{Forward (continuous line) and backward (dashed line) evolving wave functions of the photon, see Fig.2b of \cite{Zub13}.}
\label{result2}
\end{figure}

In contrast,  click in $D_2$ provides fully counterfactual information for ``logic 1" case. When Bob blocks the channel, there is no overlap of the forward and the backward evolving wave functions of the photon in the transmission channel, and the outcome of the weak measurement of the projection is zero. Thus Eve, performing weak measurements of the projection on the transmission channel can get some information about logical bits.

This work has been supported in part by grant number 32/08 of the Binational Science Foundation and the Israel Science Foundation  Grant No. 1125/10.

L. Vaidman\\
 Raymond and Beverly Sackler School of Physics and Astronomy\\
 Tel-Aviv University, Tel-Aviv 69978, Israel

\end{document}